# Magneto-elastic coupling model of deformable anisotropic superconductors

Yingxu Li [1], Guozheng Kang [1,2] and Yuanwen Gao [3,4]

1 Applied Mechanics and Structure Safety Key Laboratory of Sichuan Province, School of Mechanics and Engineering, Southwest Jiaotong University, Chengdu, Sichuan 610031, PR China
2 State Key Laboratory of Traction Power, Southwest Jiaotong University, Chengdu, Sichuan 610031, PR China
3 Key Laboratory of Mechanics on Environment and Disaster in Western China, The Ministry of Education of China, Lanzhou, Gansu 730000, PR China
4 Department of Mechanics and Engineering Science, College of Civil Engineering and Mechanics, Lanzhou University, Lanzhou, Gansu 730000, PR China

Corresponding authors: yingxuli@swjtu.edu.cn (Y. Li)

**Abstract**: We develop a magneto-elastic (ME) coupling model for the interaction between the vortex lattice and crystal elasticity. The theory extends the Kogan-Clem's anisotropic Ginzburg-Landau (GL) model to include the elasticity effect. The anisotropies in superconductivity and elasticity are simultaneously considered in the GL theory frame. We compare the field and angular dependences of the magnetization to the relevant experiments. The contribution of the ME interaction to the magnetization is comparable to the vortex-lattice energy, in materials with relatively strong pressure dependence of the critical temperature. The theory can give the appropriate slope of the field dependence of magnetization near the upper critical field. The magnetization ratio along different vortex frame axes is independent with the ME interaction. The theoretical description of the magnetization ratio is applicable only if the applied field moderately close to the upper critical field.

**Keywords**: Magneto-elastic interaction, uniaxial superconductor, anisotropic GL theory, magnetization

## 1. Introduction

It is well known that, the interaction between the defect-induced strain and the flux vortices in superconductors gives rise to the flux-pinning behaviors [1, 2]. Similarly, the elastic deformation caused by vortices, the so called magneto-elastic (ME) interaction, affects the energy of the vortex lattice. The vortex-induced strain is due to different specific volumes between the normal phase (vortex core) and the superconducting phase (superconducting matter around the vortex). The normal-state vortex cores acts as homogeneous strain sources, generating local deformations in the surrounding superconducting matter (this physical phenomenon is called the $\Delta V$ effect), such that affects the vortex-lattice energy.

The vortex-induced elastic deformation is important for the macroscopic magnetic properties. Generally speaking, the vortex-induced strain is relatively weak with respect to the pinning interaction. However, in some particular superconducting materials, the vortex-induced strain can significantly change the macroscopic magnetization performance, and may cause some anomalous magnetization [3, 4]. The critical temperature $T_c$ in these materials strongly depends on the deformation [5, 6]. In iron pnictides system, particularly Ca(Fe$_{1-x}$Co$_x$)$_2$As$_2$, the change rate of strain dependence in $T_c$ is one or two orders of magnitudes higher than conventional superconductors. This makes the ME effect outstanding in Fe-based superconductors.

The ME interaction in superconductors is a possible reason for the experimentally observed interactions between the vortex lattice and crystal lattice. The observations [7-9] of the flux-line lattice structures in NbSe$_2$ single crystal are remarkably different from the predictions with London theory. Other factor is proposed to affect the vortex interactions. However, it has not been included in the classic London theory and Ginzburg-Landau (GL) model. This factor is the ME interaction, which plays a crucial role in the formation of vortex lattice [10]. Vortices can interact one another through the elastic deformation field, and the ME interaction energy depends on the vortex-lattice structure. Some relevant experimental results of NbSe$_2$ can be described [10] by introducing the ME interaction in the London theory.

Considering the superconducting anisotropy and elasticity anisotropy in materials (as they have in nature), the ME interaction affects essentially the relationship of the vortices and crystal lattice [10]. As highly anisotropic superconductors are widely used in experiments and engineering practice, it is particularly meaningful to investigate the interaction between the vortex-lattice and crystal under considerations of anisotropic superconductivity and elasticity. For instance, the elastic interaction between vortices is of remarkable anisotropic characteristic [11]: for a set of particular elastic modulus, the two vortices located at [100] or [010] attract each other elastically; in contrary, there is repulsion between the vortices at [110]. In a relatively large magnetic field, the ME interaction removes the directional degradation of standard $60°$ triangular vortex lattice, and causes the distortion of this vortex lattice. This qualitatively consistent with the experimental data of KFe$_2$As$_2$ [12].

An effective method [10, 13-15] to quantitatively evaluate the ME interaction is introducing the vortex-induced elastic deformation in the GL theory, similar to the approach [2] of evaluating the strain field induced by the pinning defects. In the strain-dependent GL model, the $\Delta V$ effect (difference in the specific volume between normal phase and superconducting phase) is taken as the main elastic strain sources. Another theoretical model [5, 11] regards the vortices as one-dimensional strain sources in an infinite crystal, simplifying the original problem into a plane elastic problem. Taking vortex cores as the point sources of stress, the ME problem is analogy to a thermal diffusion problem in anisotropic objects subject to point heat source. This method is a London-type treatment, and only meets the requirements of qualitative calculation. A more rigorous consideration [16] of the ME interaction includes both the effects of the vortex-core region and the non-core region in the vortex-lattice energy. Although the order parameter changes slowly in the non-core region, in high-$\kappa$ superconductors the non-core region is dominant in producing elastic deformation [16]. Besides, based upon the strain-dependent GL model, some analytical solution methods [17, 18] for the ME interaction in superconductors with specific shape are proposed.

In this paper, we develop a ME model accounting for the interaction of the vortex lattice and crystal elasticity. The theory extends the Kogan-Clem's anisotropic GL model [19] to consider the ME interaction. The superconducting anisotropy and elastic anisotropy are simultaneously included in the model, in contrast to the previous theories only considering one type of anisotropy. The theoretical results are compared with some magnetization experiments, and the effects of the ME interaction as well as the theory applicability are discussed. The paper is structured as: in section 2, we establish the ME model and give the solutions near the upper critical field; in section 3, we apply the theory to calculate the elastic energy and magnetization; the theoretical results are compared with the relevant experiments in section 4; at last, we give the conclusions of the paper.

**2. Magneto-elastic interaction in anisotropic superconductors**

*2.1 Anisotropic Ginzburg-Landau equations with strain effect*

The GL theory with a phenomenological mass tensor $M_{ij}$ describes reasonably the major behaviors of anisotropic superconductors near the critical temperature $T_c$. A new feature in deformable superconductors is the vortex-induced strain in terms of the elastic response of the crystal in the presence of vortex lattices. The free energy is [10]

$$F = \int f dv = \int (\alpha|\Psi|^2 + \frac{1}{2}\beta|\Psi|^4 + \frac{H^2}{8\pi} + \frac{1}{2}M_{ij}^{-1}\Pi_i\Psi\Pi_j^*\Psi^* + \frac{1}{2}\lambda_{ijkl}u_{ij}u_{kl})dV . \qquad (1)$$

Here, $\mathbf{\Pi} = -i\hbar\nabla - (2e/c)\mathbf{A}$, $\mathbf{A}$ is the vector potential of the local magnetic field $\mathbf{H}$, $\Psi$ is the order parameter, $\lambda_{ijkl}$ are elastic coefficients and $u_{ij}$ is the strain tensor. The inverse mass tensor $M_{ij}^{-1}$ has the principal values $M_i^{-1}(i=1,2,3)$. $\alpha$ and $\beta$ are the coefficients in the GL functional expansion. In terms of the well-established strain dependence of $T_c$, the vortex-induced strain enters in the free energy via the material characteristic $\alpha$ [10]:

$$\alpha = \frac{\hbar^2}{2\bar{M}\xi_0^2 T_{c0}}(T - T_{c0}) + \alpha_{ij}u_{ij} , \qquad (2)$$

with

$$\alpha_{ij} = -\frac{\hbar^2}{2\bar{M}\xi_0^2 T_{c0}}\left(\frac{\partial T_c}{\partial u_{ij}}\right)_0 \qquad (3a)$$

[The mean mass $\bar{M} = (M_1 M_2 M_3)^{1/3}$, $T_{c0}$ is $T_c$ at zero strain and $\xi_0$ is the GL coherence length $\xi$ at temperature 0K, i.e. $\xi = \xi_0/(1-t)^{1/2}$ with $t = T/T_{c0}$].

To make the free energy dimensionless, $\Psi_0$ ($\Psi_0^2 = |\alpha_0|/\beta_0$) and $\sqrt{2}H_{c0}$ ($H_{c0}^2/4\pi = \alpha_0^2/\beta_0$) are taken as the units of the order parameter and magnetic field, respectively. As usual, one takes the average London penetration depth $\lambda_L = (\bar{M}c^2/16\pi e^2 |\Psi_0|^2)^{1/2}$ as the unit length. The units of the energy density $f$ and elastic moduli $\lambda$ are $H_{c0}^2/4\pi$. $\alpha_{ij}$ of Eq. (3) acquires the dimensionless form

$$\alpha_{ij} = -\left(\frac{\partial T_c}{\partial u_{ij}}\right)_0, \qquad (3b)$$

where the dimensionless $T_c$ has the unit $(T - T_{c0})$.

The free energy now reads

$$F = \int (-|\Psi|^2 + \frac{1}{2}|\Psi|^4 + H^2 + \frac{1}{2}\mu_{ij}^{-1}\Pi_i\Psi\Pi_j^*\Psi^* + \frac{1}{2}\lambda_{ijkl}u_{ij}u_{kl} + \alpha_{ij}u_{ij}|\Psi|^2)dV. \qquad (4)$$

Here, all quantities are dimensionless, with the same notations as their dimensional counterparts. $\Pi_i = (i\kappa)^{-1}\frac{\partial}{\partial x_i} - A_i$ where the GL parameter $\kappa = 2\sqrt{2}eH_{c0}\lambda_L^2/\hbar c$. The inverse dimensionless mass tensor $\mu_{ij} = \bar{M}M_{ij}^{-1}$ has the eigenvalues $\mu_i = \bar{M}M_i^{-1}$.

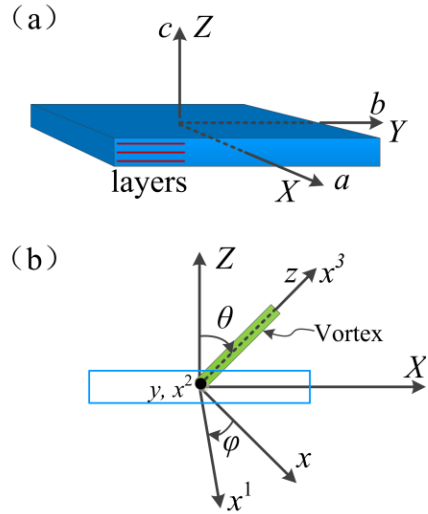

Fig. 1. (a) The crystal coordinate system $(X,Y,Z)$ coincides with the principal axes of the crystal $a$, $b$ and $c$ in a uniaxial layered superconductor. (b) The vortex coordinate frame $(x,y,z)$ is obtained by rotating $(X,Y,Z)$ with an angle $\theta$ around axis $Y$. The axis $z$ gives the vortex direction. Introducing the oblique coordinate system $(x^1, x^2, x^3)$, the anisotropic ME problem transforms into a quasi-isotropic one.

Varying the free energy (4) with respect to $\Psi^*$, $A_i$ and $\mu_{ij}$, the equations of equilibrium in the vortex frame (Fig. 1) can then be presented as [10, 16, 19]:

$$\mu_{ij}\Pi_i\Pi_j\Psi = (1 - \alpha_{ij}u_{ij})\Psi - |\Psi|^2\Psi, \qquad (5a)$$

$$e_{ikl}\frac{\partial H_l}{\partial x_k} = \mu_{ik}\operatorname{Re}(\Psi^*\Pi_k\Psi), \qquad (5b)$$

$$\lambda_{ijkl}\langle u_{kl}\rangle + \alpha_{ij}\langle |\Psi|^2\rangle = 0, \tag{5c}$$

The elasticity equation $\partial \sigma_{ij}/\partial x_j = 0$ where $\sigma_{ij} = \partial F/\partial u_{ij}$ gives

$$\frac{\partial}{\partial x_j}(\lambda_{ijkl}u_{kl} + \alpha_{ij}|\Psi|^2) = 0. \tag{5d}$$

Eqs. (5a)-(5d) describes the ME behavior of deformable superconductors in the presence of vortex-induced strain.

The solutions for the GL equations near the upper critical field $H_{c2}$ have been well established for the isotropic case [20]. Based on this consideration, we introduce a new coordinate system $(x^1, x^2, x^3)$ (Fig. 1) where the inverse mass tensor $\mu_{ij}$ has the unit matrix form, i.e. $\mu_{ij} = \delta_{ij}$ (see Appendix A for the details). In the new frame, Eqs. (5a)-(5d) read

$$\Pi_i\Pi_i\Psi = (1-\alpha^{ij}u_{ij})\Psi - |\Psi|^2\Psi, \tag{6a}$$

$$e^{ikl}\frac{\partial H_l}{\partial x^k} = \text{Re}(\Psi^*\Pi_i\Psi), \tag{6b}$$

$$\lambda^{ijkl}\langle u_{kl}\rangle + \alpha^{ij}\langle |\Psi|^2\rangle = 0, \tag{6c}$$

$$\frac{\partial}{\partial x^j}(\lambda^{ijkl}u_{kl} + \alpha^{ij}|\Psi|^2) = 0, \tag{6d}$$

*2.2 Solutions at the upper critical field $H_{c2}$*

At $H_{c2}$, the strain effect $\alpha^{ij}u_{ij}$ and high-order $|\Psi|^2$ can be neglected [19], and Eq. (6a) now reads

$$\left[(i\kappa)^{-1}\frac{\partial}{\partial x_i} - A_i\right]^2\Psi = \Psi. \tag{7}$$

This looks exactly like the isotropic case of Abrikosov's classical treatment [20]. The solutions of Eq. (7) are $H^1 = H^2 = 0$ and $H^3 = \kappa$. Note that $H^3$ is defined in the new frame $(x^1, x^2, x^3)$. While in the vortex frame $(x, y, z)$, $H_z = (\mu_1\mu_{xx})^{-1/2}H^3 = (\mu_1\mu_{xx})^{-1/2}\kappa$ [Eq. (A16)]. The upper critical field $H_{c2}$ then reads

$$H_{c2} = \tilde{\kappa} = (\mu_1\mu_{xx})^{-1/2}\kappa. \tag{8a}$$

The angular dependence $\mu_{xx} = \mu_1\cos^2\theta + \mu_3\sin^2\theta$ [Eq. (A2)] yields

$$H_{c2} = \sqrt{2}H_{c0}\tilde{\kappa} = \frac{H_{c2}^{\|ab}}{\sqrt{\sin^2\theta + \gamma^2\cos^2\theta}} = \frac{\gamma H_{c2}^{\|c}}{\sqrt{\sin^2\theta + \gamma^2\cos^2\theta}} = \frac{H_{c2}^{iso}}{\sqrt{\sin^2\theta + \gamma^2\cos^2\theta}}. \tag{8b}$$

where $\gamma^2 = \mu_1/\mu_3 = m_c/m_{ab} > 1$ for layered material, $H_{c2}^{\|ab} = H_{c2}|_{\theta=\pi/2}$, $H_{c2}^{\|c} = H_{c2}|_{\theta=0}$ and $H_{c2}^{iso} = H_{c2}|_{\gamma=1}$.

Now, we turn to Eq. (6b). If we express the order parameter as $\Psi = \sqrt{\omega}\exp(i\chi)$, then using the gauge-invariant supermomentum $Q_i$ and the current density $J^i$:

$$Q_i = \kappa^{-1}\frac{\partial \chi}{\partial x^i} - A_i, \quad J^i = e^{ikl}\frac{\partial H_l}{\partial x^k}, \tag{9}$$

Eq. (6b) is rewritten as

$$J^i = \omega Q_i. \tag{10}$$

At $H_{c2}$, the anisotropy dose not change the features of $Q_i$ in the isotropic case drastically [19]: $Q_1$ and $Q_2$ remains large, however,

$$Q_3 = 0, \quad J^3 = 0. \tag{11a}$$

Going back to the original coordinate frame $(x, y, z)$, we obtain with the help of Eq. (A16):

$$\frac{J_z}{J_x} = \frac{\mu_{xz}}{\mu_{xx}}. \tag{11b}$$

Using the angular dependence of $\mu_{xz}$ and $\mu_{xx}$ [Eq. (A2)], we have

$$\frac{J_z}{J_x} = \frac{(\mu_1 - \mu_3)\sin\theta\cos\theta}{\mu_1 \cos^2\theta + \mu_3 \sin^2\theta}. \tag{11c}$$

This reveals an important distinct feature in anisotropic superconductors at the immediate vicinity of $H_{c2}$: the axial currents do exist in an array of vortices, and these currents vanishes when the vortices direct along one of the principal crystal axes $a$ and $c$. In isotropic materials ($\mu_1 = \mu_3$), there is no axial currents at any vortex directions.

### 2.3 Solutions near $H_{c2}$: Abrikosov identities in anisotropic case with magneto-elastic effect

As in original Abrikosov approach where the well-known Abrikosov identities are derived for the isotropic case without strain effect, we introduce the operators $\Pi^{\pm} = \Pi_1 \pm i\Pi_2$ and then Eq. (6a) reads (see Appendix B for the detailed derivations)

$$\Pi^-\Pi^+\Psi = (1 - \alpha^{ij}u_{ij} - \kappa^{-1}H^3 - |\Psi|^2 - Q_3^2)\Psi. \tag{12}$$

Accounting for Eq. (11a), $Q_3$ has the order of $(H_{c2} - H)/H_{c2}$ near $H_{c2}$. All terms on the RHS (right hand side) of Eq. (12) can be neglected in the linear approximation. Instead of solving Eq. (12), we have a more simplified form, $\Pi^-\Pi^+\Psi = 0$, which is further simplified to

$$\Pi^+\Psi_0 = 0. \tag{13}$$

Using $\Pi^+ = \Pi_1 + i\Pi_2$, $\Psi_0 = \sqrt{\omega}\exp(i\chi)$ and $Q_i = \kappa^{-1}\frac{\partial \chi}{\partial x^i} - A_i$ in Eq. (13), we have $\omega Q_1 = -\frac{1}{2\kappa}\frac{\partial \omega}{\partial x^2}$ and $\omega Q_2 = \frac{1}{2\kappa}\frac{\partial \omega}{\partial x^1}$. Combining this with $\omega Q_1 = \frac{\partial H_3}{\partial x^2}$ and $\omega Q_2 = -\frac{\partial H_3}{\partial x^1}$ [Eqs. (9) and (10)], we obtain the first Abrikosov identity in anisotropic case:

$$H_3 = \text{const} - \frac{\omega}{2\kappa}, \tag{14a}$$

or in the vortex frame,

$$H_z = H_0 - \frac{\omega}{2\tilde{\kappa}}, \tag{14b}$$

where $H_0$ is an arbitrary constant, and $\tilde{\kappa}$ has been given in Eq. (8a).

To normalize the solution $\Psi_0$ of the homogeneous equation (13), one has to find the solution $\Psi_0 + \Psi_1$ to the exact nonlinear equation (12). Substituting $\Psi = \Psi_0 + \Psi_1$ and $H^3 = (\mu_1 \mu_{xx})^{1/2}(H_z - \mu_{xz}\mu_{xx}^{-1}H_x)$ [see Eq. (A16)] in Eq. (12), and with the help of Eqs. (13) and (14b), we have

$$\Pi^-\Pi^+\psi_1 = \psi_0 [\frac{\tilde{\kappa} - H_0}{\tilde{\kappa}} - \alpha^{ij}u_{ij} + (\frac{1}{2\tilde{\kappa}^2} - 1)\omega + 2\bar{\gamma}H_x]. \tag{15}$$

where $\bar{\gamma} = \frac{\mu_{xz}}{2\mu_{xx}\tilde{\kappa}}$. We have omitted $Q_3^2$ in the derivation of Eq. (15), since it is of higher order of $(H_{c2} - H)/H_{c2}$ near $H_{c2}$. The existence of a solution $\psi_1$ for the inhomogeneous linear equation (15) requires that, the RHS of Eq. (15) is orthogonal to the solution $\Psi_0$ of the corresponding homogenous equation $\Pi^+\Psi_0 = 0$ [19]. This leads to

$$\frac{\tilde{\kappa} - H_0}{\tilde{\kappa}} \langle \omega \rangle + \left(\frac{1}{2\tilde{\kappa}^2} - 1\right)\langle \omega^2 \rangle + 2\bar{\gamma}\langle H_x \omega \rangle - \alpha^{ij}\langle u_{ij}\omega \rangle = 0, \tag{16}$$

the second Abrikosov identity. Eq. (16) generalizes the Kogan-Clem's result without the ME interaction [19], i.e. $\alpha^{ij} = 0$, to the anisotropic material with the ME effect.

*2.4 Free energy*

Bearing in mind that from Eq. (14b) the magnetic induction is

$$B_z = \langle H_z \rangle = H_0 - \frac{\langle \omega \rangle}{2\tilde{\kappa}}, \tag{17}$$

$\langle \omega \rangle$ can be obtained in terms of Eq. (16):

$$\langle \omega \rangle = \frac{2\tilde{\kappa}(\tilde{\kappa} - B_z)}{1 + (2\tilde{\kappa}^2 - 1)\frac{\langle \omega^2 \rangle}{\langle \omega \rangle^2} - 2\tilde{\kappa}^2 \frac{2\bar{\gamma}\langle H_x \omega \rangle}{\langle \omega \rangle^2} + 2\tilde{\kappa}^2 \frac{\alpha^{ij}\langle u_{ij}\omega \rangle}{\langle \omega \rangle^2}}. \tag{18}$$

If we introduce the following notations

$$\beta_A = \frac{\langle \omega^2 \rangle}{\langle \omega \rangle^2}, \quad \tilde{\beta}_A = 1 + (2\tilde{\kappa}^2 - 1)\beta_A, \quad \beta_e = -2\tilde{\kappa}^2 \frac{\alpha^{ij}\langle u_{ij}\omega \rangle}{\langle \omega \rangle^2},$$

$$\bar{\gamma}\langle H_x \omega \rangle = \langle H_\perp^2 \rangle = \bar{\gamma}^2 \frac{\beta_1(\beta_A - 1)\langle \omega \rangle^2}{2}, \quad \delta = -2\tilde{\kappa}^2 \bar{\gamma}^2 \beta_1(\beta_A - 1), \tag{19}$$

(the last one is derived from the lattice structure [19] with $H_\perp^2 = H_x^2 + H_y^2$ and $\beta_1$ the structure parameter), Eq. (18) reads

$$\langle \omega \rangle = \frac{2\tilde{\kappa}(\tilde{\kappa} - B_z)}{\tilde{\beta}_A - \beta_e + \delta}. \tag{20}$$

The general form of the free energy in anisotropic superconductors with elasticity effect is [16, 19]

$$F = \left\langle H_z^2 + H_\perp^2 - \frac{\omega^2}{2} + \frac{1}{2}\lambda_{ijkl}u_{ij}u_{kl} \right\rangle. \tag{21}$$

Using Eqs. (14b) (17) and (19) in Eq. (21), the free energy (21) is rewritten as

$$F = B_z^2 - \langle \omega \rangle^2 \frac{\tilde{\beta}_A + \delta}{4\tilde{\kappa}^2} + \frac{1}{2}\lambda^{ijkl}\langle u_{ij}u_{kl}\rangle, \tag{22a}$$

and expressing $\langle \omega \rangle$ in terms of Eq. (20), one has another form of the free energy:

$$F = B_z^2 - \frac{(\tilde{\kappa} - B_z)^2(\tilde{\beta}_A + \delta)}{(\tilde{\beta}_A - \beta_e + \delta)^2} + \frac{1}{2}\lambda^{ijkl}\langle u_{ij}u_{kl}\rangle. \tag{22b}$$

Now let us go further with the last term in the free energy (22a), considering the effect of the vortex-induced strain with the help of elasticity equations (6c) and (6d). The Fourier form of the vortex induced strain $u_{ij}$ reads [16]

$$u_{ij} = \bar{u}_{ij} + \frac{i}{2}\sum_{q\neq 0}[q_i u_j(\mathbf{q}) + q_j u_i(\mathbf{q})]e^{i\mathbf{q}\cdot\mathbf{p}}, \tag{23}$$

where $u_i(\mathbf{q})$ is the displacement component in Fourier space, and $\bar{u}_{ij}$ is the homogenous strain induced by the vortex. Applying Eq. (23) in Eqs. (6c) and (6d), we have

$$\lambda^{ijkl}\bar{u}_{ij} + \alpha^{ij}\langle \omega \rangle = 0, \tag{24a}$$

$$G_{ij}^{-1}(\mathbf{q})u_j(\mathbf{q}) - iS_i(\mathbf{q})\omega(\mathbf{q}) = 0, \tag{24b}$$

where $G_{ik}^{-1}(\mathbf{q}) = \lambda^{ijkl}q_j q_l$, $S^i(\mathbf{q}) = \alpha^{ij}q_j$ and $\omega(\mathbf{q})$ is the Fourier transform of Eq. (20). Solving Eq. (24a) and (24b), we have the strain field in deformable superconductors:

$$\bar{u}_{ij} = -\frac{\alpha^{kl}\langle \omega \rangle}{\lambda^{ijkl}}, \tag{25a}$$

$$u_i(\mathbf{q}) = iS^j(\mathbf{q})G_{ji}(\mathbf{q})\omega(\mathbf{q}). \tag{25b}$$

With the help of Eqs. (25a) and (25b), the last term in the free energy (22a) now can be expressed as (the detailed derivation can be found in [16])

$$\frac{1}{2}\lambda^{ijkl}\langle u_{ij}u_{kl}\rangle = \frac{1}{2}\left[\lambda^{ijkl}\bar{u}_{ij}\bar{u}_{kl} + \sum_{q\neq 0}G_{ij}^{-1}(\mathbf{q})u_i(\mathbf{q})u_j(-\mathbf{q})\right] = -\frac{1}{2}\alpha^{ij}\langle u_{ij}\omega \rangle = \frac{\beta_e}{4\tilde{\kappa}^2}\langle \omega \rangle^2. \tag{26}$$

The final form of the free energy can be written as

$$F = B_z^2 - \frac{\tilde{\beta}_A - \beta_e + \delta}{4\tilde{\kappa}^2}\langle \omega \rangle^2, \tag{27a}$$

or with the help of Eq. (20),

$$F = B_z^2 - \frac{(\tilde{\kappa} - B_z)^2}{\tilde{\beta}_A - \beta_e + \delta}. \tag{27b}$$

*2.5 Comparisons with the results in the classical theories*

Let us validate Eq. (20) by comparing it to the classical results. If the superconducting material is isotropic but with strain dependence, i.e. $\bar{\gamma} = \delta = 0$, $\tilde{\kappa} = \kappa$ and $\beta_e \neq 0$, Eq. (20) reduces to the result $\langle\omega\rangle = \dfrac{2\kappa(\kappa - B)}{\tilde{\beta}_A - \beta_e}$ by Cano et al. [16]. On the other hand, if the superconducting material is anisotropic but without strain dependence, i.e. $\beta_e = 0$, one obtains from Eq. (20) that $\langle\omega\rangle = \dfrac{2\tilde{\kappa}(\tilde{\kappa} - B_z)}{\tilde{\beta}_A + \delta}$. The last one coincides with Kogan-Clem's result [19].

The validity of Eq. (27) can be checked by comparing it to the classical results: in the particular case where the superconducting anisotropy vanishes, i.e. $\delta = 0$, $\tilde{\kappa} = \kappa$ and $B_z = B$, Eq. (27b) reduces to $F = B^2 - \dfrac{(\kappa - B)^2}{\tilde{\beta}_A - \beta_e}$. The last one just reproduces the result by Cano et al. [16]. If the material superconductivity is independent with strain, i.e. $\beta_e = 0$, Eq. (27b) takes the form of $F = B_z^2 - \dfrac{(\tilde{\kappa} - B_z)^2}{\tilde{\beta}_A + \delta}$, which coincides with the Kogan-Clem's result [19].

## 3. Application of the theory

*3.1 Parameters $\beta_A$, $\beta_e$ and $\delta$ in Free energy*

Let us first determine $\delta$ and $\beta_A$ in the free energy (27b). From Eq. (19) we have $\langle H_\perp^2\rangle/\langle\omega\rangle^2 = \gamma^2 \dfrac{\beta_1(\beta_A - 1)}{2}$ which is of order of $\tilde{\kappa}^{-2}(\beta_A - 1)$ [10], so that for high-$\kappa$ superconductors $\langle H_\perp^2\rangle/\langle\omega\rangle^2 \ll \beta_A$ [10]. Thus, we can neglect $\delta$ in the free energy (27b). $\beta_A = 1.16$ is a geometric constant for the triangular lattice [20].

In deformable superconductors, the free energy (27b) depends on the strains via $\beta_e = -2\tilde{\kappa}^2 \alpha^{ij} \langle u_{ij}\omega\rangle/\langle\omega\rangle^2$ with $\alpha^{ij}$ given in Eq. (3b). For the uniform deformation, using Eq. (25a) we have $\beta_e = 2\tilde{\kappa}^2 \dfrac{\alpha^{ij}\alpha^{kl}}{\lambda^{ijkl}} \beta_A$. Now, to calculate $\beta_e$ one needs an explicit form of $\alpha^{ij}$. Regarding Eq. (3b), note that $(\partial T_c/\partial u_{ij})_0$ is always measured by the stress dependences of $T_c$. Thus, using $p^{ij} = \lambda^{ijkl} u_{kl}$, we have $\left(\dfrac{\partial T_c}{\partial u_{ij}}\right)_0 = \lambda^{ijkl} \dfrac{\partial T_c}{\partial p^{ij}}$. Finally, we express $\beta_e$ as

$$\beta_e = 2\tilde{\kappa}^2 \beta_A \dfrac{\upsilon^{ij}\upsilon^{kl}}{\lambda^{ijkl}}, \quad (28)$$

where

$$\upsilon^{ij} = \lambda^{ijmn} \dfrac{\partial T_c}{\partial p^{mn}}. \quad (29a)$$

Since $\upsilon^{ij}\upsilon^{kl}/\lambda^{ijkl}$ is an invariant independent with coordinates transformation, we can calculate $\upsilon^{ij}$ and $\lambda^{ijkl}$ [in the new frame $(x^1, x^2, x^3)$] in terms of the crystal frame $(a,b,c)$. $p^{mn}$ is the uniform stress applying on the superconductor, such that the nonzero components are $p^{11} = p^{22} = p_{ab}$ and $p^{33} = p_c$. Hereafter, we consider the elastic medium as tetragonal crystal lattice. In the crystal frame, the nonzero components of the elastic modulus tensor $\lambda_{ijkl}$ [in the Cartesian coordinate system $(a,b,c)$ the covariant components are equal to the corresponding contravariant ones, i.e. $\lambda^{ijkl} = \lambda_{ijkl}$] are [11]

$$\lambda_{aaaa} = \lambda_{bbbb} = C_{11}, \ \lambda_{aabb} = C_{12}, \ \lambda_{abab} = C_{66}, \ \lambda_{cccc} = C_{33}, \ \lambda_{aacc} = \lambda_{bbcc} = C_{13}, \ \lambda_{acac} = \lambda_{bcbc} = C_{55}.$$

Thus, the nonzero components of $\upsilon^{ij}$ read (bearing in mind that $\lambda_{ijkl}$ has the general symmetry properties $\lambda_{ijkl} = \lambda_{jikl} = \lambda_{ijlk} = \lambda_{lkij}$ due to the symmetry in the strain tensor)

$$\upsilon_{aa} = \upsilon_{bb} = \upsilon_{ab} = (C_{11} + C_{12})\frac{\partial T_c}{\partial p_{ab}} + C_{13}\frac{\partial T_c}{\partial p_c},$$

$$\upsilon_{cc} = \upsilon_c = 2C_{13}\frac{\partial T_c}{\partial p_{ab}} + C_{33}\frac{\partial T_c}{\partial p_c}, \tag{29b}$$

where $C_{ij}$ are the elastic moduli in the crystal frame. Thus, we have

$$\frac{\upsilon^{ij}\upsilon^{kl}}{\lambda^{ijkl}} = 2\upsilon_{ab}^2\left(\frac{1}{C_{11}} + \frac{1}{C_{12}}\right) + 4\upsilon_{ab}\upsilon_c\frac{1}{C_{13}} + \upsilon_c^2\frac{1}{C_{33}}. \tag{30a}$$

or in an estimation for the order of magnitude,

$$\frac{\upsilon^{ij}\upsilon^{kl}}{\lambda^{ijkl}} \sim \bar{\lambda}\left(\frac{\partial T_c}{\partial p}\right)^2. \tag{30b}$$

where $\bar{\lambda}$ is of the order of the elastic modulus.

Combining Eq. (28) with Eq. (30) determines $\beta_e$. So far, the parameters $\beta_A$, $\delta$ and $\beta_e$ in the free energy have been determined.

*3.2 Elastic energy*

The elastic energy can be evaluated with the help of Eqs. (20) (26) (28) and (30b):

$$F_{el} = \frac{1}{2}\lambda^{ijkl}\langle u_{ij}u_{kl}\rangle = \frac{\beta_e}{\left[1 + (2\tilde{\kappa}^2 - 1)\beta_A - \beta_e\right]^2}(\tilde{\kappa} - B_z)^2. \tag{31}$$

Here, $\beta_e = 2\tilde{\kappa}^2\beta_A\bar{\lambda}\left(\frac{\partial T_c}{\partial p}\right)^2$.

*3.3 Magnetization*

Following the approaches in [19], one needs the expressions of the field **H** to obtain the magnetization $\mathbf{M} = (\mathbf{B} - \mathbf{H})/4\pi$. In the vortex frame $(x, y, z)$, the field components $H_z$ and $H_x$ are calculated by [19]

$$H_{Mx} = -\frac{\tilde{\kappa} - B}{(\tilde{\beta}_A - \beta_e)\tilde{\kappa}} \frac{\partial \tilde{\kappa}}{\partial \theta} = -\frac{\mu_{xz}}{\mu_{xx}} \frac{\tilde{\kappa} - B}{\tilde{\beta}_A - \beta_e}, \tag{32a}$$

$$H_{Mz} = \frac{1}{2}\left(\frac{\partial F}{\partial B_z}\right)_\theta = B_z + \frac{\tilde{\kappa} - B_z}{\tilde{\beta}_A - \beta_e}. \tag{32b}$$

Now the magnetization **M** in the vortex frame reads

$$-4\pi M_x = -\frac{\mu_{xz}}{\mu_{xx}} \frac{H_{c2} - B}{\tilde{\beta}_A - \beta_e}, \tag{33a}$$

$$-4\pi M_z = \frac{H_{c2} - B}{\tilde{\beta}_A - \beta_e}, \tag{33b}$$

in conventional units. The ratio

$$\frac{M_x}{M_z} = -\frac{\mu_{xz}}{\mu_{xx}} = -\frac{(\gamma^2 - 1)\sin\theta\cos\theta}{\gamma^2 \cos^2\theta + \sin^2\theta}. \tag{34}$$

is independent with the vortex structure parameter $\tilde{\beta}_A$ and elasticity parameter $\beta_e$. This extend the conclusion by Kogan and Clem [19] to the situation with the ME effect: the magnetization ratio is independent with the ME interaction.

The magnetization components in the crystal frame $(X, Y, Z)$ is obtained by using the coordinate transformation relations in Eqs. (33a) and (33b):

$$-4\pi M_X = \frac{H_{c2} - B}{\tilde{\beta}_A - \beta_e} \frac{\mu_3}{\mu_{xx}} \sin\theta = \frac{H_{c2} - B}{\tilde{\beta}_A - \beta_e} \frac{\mu_3 \sin\theta}{\mu_1 \cos^2\theta + \mu_3 \sin^2\theta}, \tag{35a}$$

$$-4\pi M_Z = \frac{H_{c2} - B}{\tilde{\beta}_A - \beta_e} \frac{\mu_1}{\mu_{xx}} \cos\theta = \frac{H_{c2} - B}{\tilde{\beta}_A - \beta_e} \frac{\mu_1 \cos\theta}{\mu_1 \cos^2\theta + \mu_3 \sin^2\theta}. \tag{35b}$$

The magnetization ratio in the crystal frame is

$$\frac{M_X}{M_Z} = \frac{\mu_3}{\mu_1} \tan\theta = \gamma^{-2} \tan\theta. \tag{36}$$

## 4. Comparisons with experiments

*4.1 Angular dependence of the upper critical field*

Ghosh et al. [21] measure the anisotropic upper critical field in CaAlSi. The experimental results can be fitted with Eq. (8b), by choosing the appropriate anisotropic parameter $\gamma$ (Fig. 2). The increases in the anisotropy parameter $\gamma$ with temperature coincides with the measurements in the same experiment. The similar temperature dependence is found in LaFeAsO$_{1-x}$F$_x$ thin films [22], ranging from 3.2 at 2 K to 4.2 at 15 K. However, for MgB$_2$, $\gamma$ is decreasing with temperature [23]. The electronic band structure and anisotropic nature of the order parameter are responsible for these behaviors.

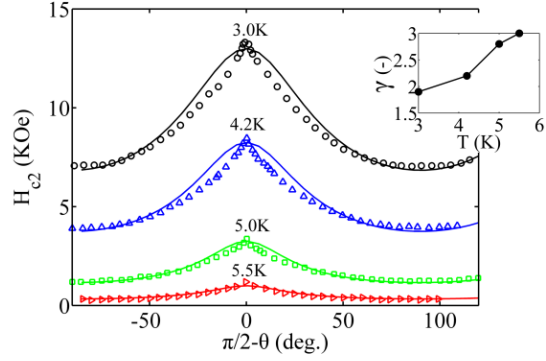

Fig. 2. The upper critical field as a function of the angle $\pi/2-\theta$ at temperatures 3.0, 4.2, 5.0 and 5.5 K. Eq. (8b) (lines) fits roughly well the measured $H_{c2}(\theta)$ (dots) in CaAlSi single crystal [21]. The insert shows the temperature dependence of the anisotropy parameter $\gamma$, which is taken as the fitting parameter in Eq. (8b).

*4.2 Field dependence of mmagnetization*

The theory considers the mediate vicinity of $B_{c2}$, and high-$\kappa$ material, such that from Eq. (32) the applied field $H_M$ approaches the magnetic induction $B_z$, i.e. $H_M \approx H_{Mz} \approx B_z$ and $H_{Mx} \approx 0$. The parameters in Eq. (33) can be rewritten as (in conventional units)

$$\tilde{\beta}_A = 1 + (2\tilde{\kappa}^2 - 1)\beta_A, \tag{37a}$$

$$\beta_e = \frac{\tilde{\kappa}^2 \beta_A \overline{\lambda} H_{c0}(0)^2}{2\pi T_c^2}\left(\frac{\partial T_c}{\partial p}\right)^2, \tag{37b}$$

$$\tilde{\kappa} = \frac{H_{c2}^{\|ab}}{\sqrt{2}H_{c0}}\frac{1}{\sqrt{\sin^2\theta + \gamma^2\cos^2\theta}} = \frac{H_{c2}^{\|c}}{\sqrt{2}H_{c0}}\frac{\gamma}{\sqrt{\sin^2\theta + \gamma^2\cos^2\theta}}, \tag{37c}$$

for calculation convenience.

According to the magnetization experiment [24] in La$_{1.45}$Nd$_{0.40}$Sr$_{0.15}$CuO$_4$ [24], we take $T_c = 10.5$ K, $H_{c0}(6K) = 0.11$ T, $H_{c0}(8K) = 0.05$ T, $H_{c0}(0) = 0.2$ T, $H_{c2}^{\|c}(6K) = 4.8$ T, $H_{c2}^{\|c}(8K) = 4.5$ T and $\theta = 0$ in Eq. (33). Since the mechanical parameters are not involved in this experiment, we take $\partial T_c/\partial p = 3\times 10^{-9}$ K cm$^3$/erg and $\overline{\lambda} = 1\times 10^{12}$ erg/cm$^3$ according to a similar crystal La$_{1.45}$Nd$_{0.40}$Sr$_{0.14}$CuO$_4$ as suggested by [5]. We obtain $M(H)$ shown in Fig. 3(a). The calculation curve describes well the experimental data near $H_{c2}$, but there are relatively large deviations away from $H_{c2}$. This is very important since the theory is essentially established at the intermediate vicinity of $H_{c2}$. The contributions to the slope $dM_z/dB = \left[4\pi\left(\tilde{\beta}_A - \beta_e\right)\right]^{-1}$ include the magnetic part $\tilde{\beta}_A$ and the ME interaction $\beta_e$. Since the superconductivity is sensitive to pressure in $La_{1.45}Nd_{0.40}Sr_{0.15}CuO_4$ [5], i.e. considerable $\partial T_c/\partial p$ value, $\beta_e = 4.3\times10^3$ is comparable to $\tilde{\beta}_A = 9.4\times10^3$. Without including the elasticity effect, i.e. setting $\beta_e = 0$ in Eq. (33b), the agreement around $H_{c2}$ is relatively poor [Fig. 3(b)], since the slope of $M(H)$ is not properly evaluated.

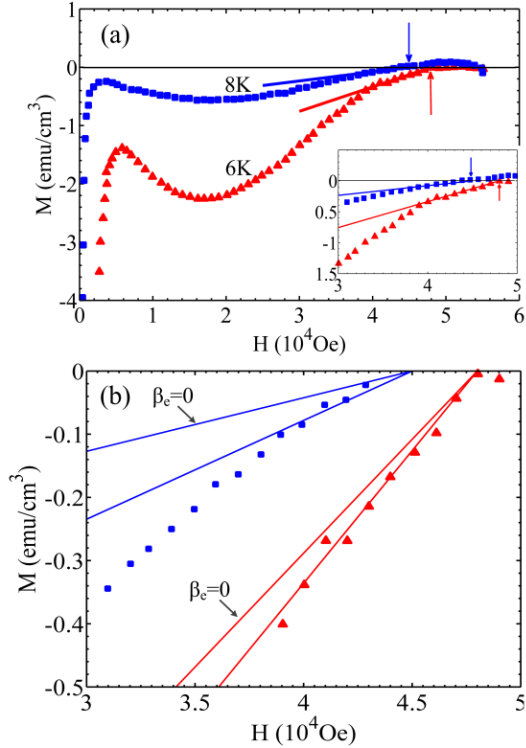

Fig. 3. (a) The field dependence of the magnetization. The lines are obtained with Eq. (33b), to compare the experimental data in $La_{1.45}Nd_{0.40}Sr_{0.15}CuO_4$ [24]. The arrows indicate the upper critical field $H_{c2}$, where the magnetization increases to 0. The insert shows the magnification near $H_{c2}$. (b) Comparison with the result of $\beta_e = 0$, i.e. the case without elasticity effect.

*4.3 Angular dependence of magnetization ratio*

To check the anisotropic magnetization ratio Eq. (34), we fit Eq. (34) to the experimental magnetization data in $YBa_2Cu_3O_{7-x}$ at 70K and 5T [25]. The fit gives the anisotropic parameter $\gamma = 4.5$ [Fig. 4(a)]. This is consistent with the reported values 3~10 in other literatures [26-28]. Taking $H_{c2}^{\|c}(70K) = 6$ T (roughly corresponding to the experimental value in [29]) and $\gamma = 5.5$, we obtain $H_{c2}(\theta)$ for this case. Within the range $60° < \theta < 90°$, the fit is relatively poor. This can be attributed to the narrow peak range of $H_{c2}(\theta)$, where the $H_{c2}$ values are several times larger than the applied field $H = 5T$. The latter deviate severely from the applicability range $|H_{c2} - H| \ll H_{c2}$.

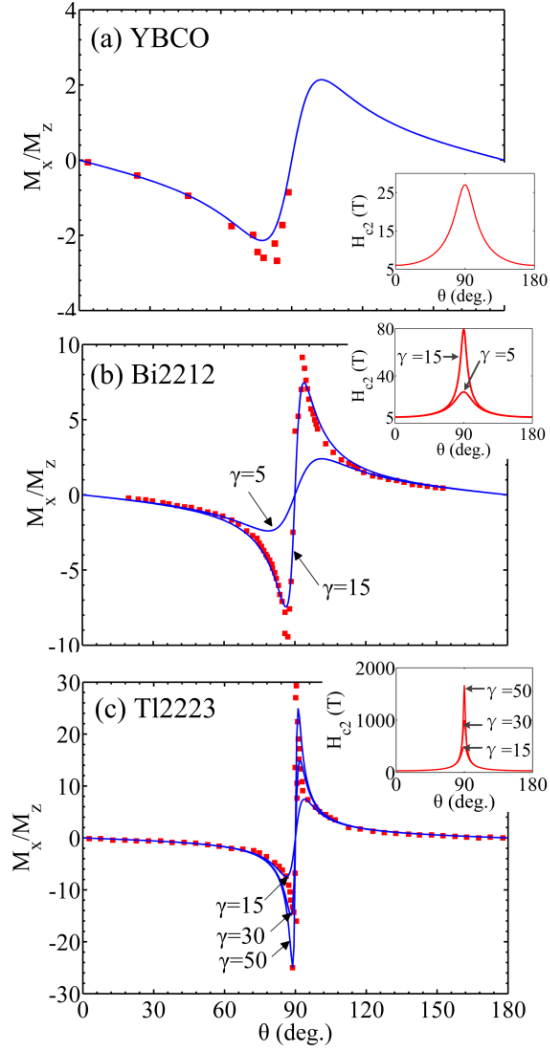

Fig. 4. (a) Angular dependence of the $M_x/M_z$ ratio in YBCO crystal. The experimental data [25] (solid square) are measured at temperature 70K and magnetic field 5T. The line corresponds to Eq. (34) using $\gamma = 5$. The insert shows the $H_{c2}(\theta)$ at this $\gamma$ value. (b) Similar with (a), but for Bi-2212 crystal at 70K and 1T. The lines correspond to Eq. (34) with different $\gamma$ values. (c) *ibid,* but for Tl-2223 crystal at 110K and 1T.

We also fit Eq. (34) to the $M_x/M_z$ data for Bi-2212 crystal [25] at $H = 1$T and $T = 70$K, by varying $\gamma$ to obtain the best fit [Fig. 4(b)]. The fitting curve of $\gamma = 5$ reproduce fairly the experimental data in a wide range except for $60° < \theta < 120°$. $H_{c2}(\theta)$ is obtained by taking $H_{c2}^{\|c}(70\text{K}) = 5.3$ T [30] and $\gamma = 5$ in Eq. (8b). Compared with the good-fitting range, the poor-fitting range shows several times higher values of $H_{c2}$. Similar with YBCO, in this case, the difference between the applied field $H$ and $H_{c2}$ determines the fitting accuracy. We also take $\gamma = 15$ to fit the data; this value approaches $\gamma \approx 17$ proposed by Tuominen *et al.* [25]. This basically resolves the deviations in the region $60° < \theta < 120°$. However, $H_{c2}$ is nearly two orders of magnitude higher than the applied field. $|H_{c2} - H|$ for $\gamma = 15$ are seriously larger than $\gamma = 5$. It must be noted, the seemingly good fit by $\gamma = 15$ is actually out of scope of the theory, so it masks the physics of the anisotropic magnetization.

We implement the fit for the Tl-2223 samples, which is supposed to have remarkably high $\gamma$ and $H_{c2}$ values, i.e. $\gamma \geq 200$ [31] and $H_{c2}^{\parallel c}(110K) = 33.4T$ [32]. In this case, the agreement is fairly good for $|\theta - 90°| \geq 10°$, independently of the $\gamma$ value used above ~15 [Fig. 4(c)]. With higher $\gamma$ value, i.e. $\gamma = 30$ and 50, the amplitude of the peak approaches the experimental data at the vicinity of $90°$. However, the high $\gamma$ values lead to the $H_{c2}$ two or three orders of magnitude higher than the applied field $H = 1$ in this region. So, the theory is not applicable in this case, even for the small-$\theta$ regime. This conclusion can also be verified by comparing Eq. (36) with the experimental [31] magnetization ratio along the crystal axes. Equation (36) establishes the angular dependence of $M_X/M_Z$: a larger $\theta$ corresponds to a higher ratio value. In contrary, however, the experiment [31] for Tl-2223 shows an opposite dependence.

## 5. Concluding remarks

Based on the anisotropic GL theory, we model the ME interaction in a uniaxial superconductor. The present model includes the anisotropic superconductivity and elastic deformation, by introducing the anisotropic electron mass and the elastic energy in the GL free energy. The GL equations are now correlated to the ME interaction. The solutions are obtained by using the anisotropic Abrikosov identities under the considerations of the elasticity effect. As a validation, the formulas of the free energy is compared with the classical results.

The elasticity parameter $\beta_e$, vortex-lattice parameter $\beta_A$ and anisotropy parameter $\delta$ are determined for practice. We derive the practical forms of free energy and magnetization. The formula of the anisotropic magnetization ratio remains its classical representation, even taking into account the ME coupling effect.

We compare the theory with the experiments. By including the ME interaction, the model gives a satisfied description of the field dependence of the magnetization near the upper critical field. In Fe-based superconductors, the strong dependence $\partial T_c / \partial p$ leads to a remarkable ME interaction, which is comparable to the vortex energy. The ME coupling effect is important for an appropriate description of the magnetization behaviors.

The effect of the ME interaction is absent in the magnetization ratio $M_x/M_z$ along different vortex frame axes. This enables the fitting of the relevant experiments to determine the anisotropic parameter $\gamma$ substantially. However, it should be noted that, excessively high $\gamma$ values lead to unrealistically large $H_{c2}$, especially when the field directs approximately in the superconducting layer ($ab$ plane). One must bear in mind the applicability range of the theory: $|H_{c2} - H| \ll H_{c2}$. The appropriate relaxation of $|H_{c2} - H| \ll H_{c2}$ considering the practical application is: $H$ and $H_{c2}$ are in the same order of magnitude.

The theory is convincible only at the vicinity of $H_{c2}$. Extending the theory to a moderate field range is significant for practice. However, in this case, the precise vortex structure and spatial distribution of order parameter should be considered. This makes the problem intricate, not to mention the anisotropies of elasticity and superconductivity. The anisotropic ME interaction in the moderate magnetic field range will be the focus of our next research.

**Acknowledgments**

This work was performed with supports from the National Natural Science Foundation of China (11372120, 11421062, 11102171 and 11302119) and National Key Project of Magneto-Restriction Fusion Energy Development Program (2013GB110002). The Fundamental Research Funds for the Central Universities (A0920502051619-95) and China Postdoctoral Science Foundation (2016M602707) are also acknowledged.

**Appendix A: Transformation of coordinates**

Let us consider a uniaxial layered superconductor (Fig. 1). In the crystal coordinate system $(X,Y,Z)$ where $Z$ axis is normal to the layers, and the components $\mu^{KL}$ $(K,L=1,2,3)$ of the inverse dimensionless mass tensor

$$\mu^{KL} = \begin{bmatrix} \mu^{11} & 0 & 0 \\ 0 & \mu^{22} & 0 \\ 0 & 0 & \mu^{33} \end{bmatrix} = \begin{bmatrix} \mu_1 & 0 & 0 \\ 0 & \mu_1 & 0 \\ 0 & 0 & \mu_3 \end{bmatrix}. \tag{A1}$$

The upper indices dictate the contravariant quantities. Supposing an array of vortices tilted from the crystal axes, we introduce a new coordinates $(x,y,z)$ of vortices. $(x,y,z)$ is rotated with respect to the $(X,Y,Z)$ through an angle $\theta$ about the $Y$ axis (Fig. 1). When $(X,Y,Z)$ transforms into $(x,y,z)$, the tensor components transform as the coordinates transformation relations $\bar{x}^1 = X^1\cos\theta - X^3\sin\theta$, $\bar{x}^2 = X^2$, $\bar{x}^3 = X^1\sin\theta + X^3\cos\theta$ (introducing the notations $X^1 = X$, $X^2 = Y$, $X^3 = Z$, $\bar{x}^1 = x$, $\bar{x}^2 = y$ and $\bar{x}^3 = z$ for the convenience of tensor analysis):

$$\begin{aligned}
\left[\mu^{\bar{i}\bar{j}}\right] &= \left[\beta_K^{\bar{i}}\beta_L^{\bar{j}}\mu^{KL}\right] = \begin{bmatrix} \cos\theta & 0 & -\sin\theta \\ 0 & 1 & 0 \\ \sin\theta & 0 & \cos\theta \end{bmatrix} \begin{bmatrix} \mu_1 & 0 & 0 \\ 0 & \mu_2 & 0 \\ 0 & 0 & \mu_3 \end{bmatrix} \begin{bmatrix} \cos\theta & 0 & \sin\theta \\ 0 & 1 & 0 \\ -\sin\theta & 0 & \cos\theta \end{bmatrix} \\
&= \begin{bmatrix} \mu_1\cos^2\theta + \mu_3\sin^2\theta & 0 & (\mu_1-\mu_3)\sin\theta\cos\theta \\ 0 & \mu_2 & 0 \\ (\mu_1-\mu_3)\sin\theta\cos\theta & 0 & \mu_1\sin^2\theta + \mu_3\cos^2\theta \end{bmatrix} \\
&= \begin{bmatrix} \mu_{xx} & 0 & \mu_{xz} \\ 0 & \mu_2 & 0 \\ \mu_{xz} & 0 & \mu_{zz} \end{bmatrix}
\end{aligned} \tag{A2}$$

where the transformation coefficients

$$\beta_K^{\bar{i}} = \frac{\partial \bar{x}^i}{\partial X^K} = \begin{bmatrix} \cos\theta & 0 & -\sin\theta \\ 0 & 1 & 0 \\ \sin\theta & 0 & \cos\theta \end{bmatrix}. \tag{A3}$$

The three invariants of the tensor, $\text{tr}(\boldsymbol{\mu})$, $\boldsymbol{\mu}\cdot\cdot\boldsymbol{\mu}$ and $\det(\boldsymbol{\mu})$ give the following useful relations

$$\mu_{xx} + \mu_{zz} = \mu_1 + \mu_3, \quad \mu_{xx}\mu_{zz} - \mu_{xz}^2 = \mu_1\mu_3. \tag{A4}$$

To render $\mu^{\bar{i}\bar{j}}$ isotropic ($\mu^{\bar{i}\bar{j}} = \delta^{ij}$), we introduce an oblique angled rectilinear coordinate system $(x^1, x^2, x^3)$. The coordinate transformation between $(x, y, z)$ and $(x^1, x^2, x^3)$ is chosen as [19]

$$x^1 = a\bar{x}^1, \quad x^2 = b\bar{x}^2, \quad x^3 = c\bar{x}^1 + d\bar{x}^3. \tag{A5}$$

The inverse transformation is

$$\bar{x}^1 = a^{-1}x^1, \quad \bar{x}^2 = b^{-1}x^2, \quad \bar{x}^3 = d^{-1}x^3 - c(ad)^{-1}x^1. \tag{A6}$$

The covariant $\beta_i^{\bar{j}}$ and contravariant $\beta_{\bar{j}}^i$ transformation coefficients now read

$$\left[\beta_j^{\bar{i}}\right] = \left[\frac{\partial \bar{x}^i}{\partial x^j}\right] = \begin{bmatrix} a^{-1} & 0 & 0 \\ 0 & b^{-1} & 0 \\ -c(ad)^{-1} & 0 & d^{-1} \end{bmatrix}, \quad \left[\beta_{\bar{j}}^i\right] = \left[\frac{\partial x^i}{\partial \bar{x}^j}\right] = \begin{bmatrix} a & 0 & 0 \\ 0 & b & 0 \\ c & 0 & d \end{bmatrix}. \tag{A7}$$

Setting the covariant base vectors of $(\bar{x}^1, \bar{x}^2, \bar{x}^3)$ as $\mathbf{g}_{\bar{1}} = \mathbf{i}$, $\mathbf{g}_{\bar{2}} = \mathbf{j}$ and $\mathbf{g}_{\bar{3}} = \mathbf{k}$, one can determine the covariant base vectors $\mathbf{g}_i$ in $(x^1, x^2, x^3)$:

$$\left[\mathbf{g}_i\right] = \left[\beta_i^{\bar{j}}\mathbf{g}_{\bar{j}}\right] = \begin{bmatrix} a^{-1} & 0 & -c(ad)^{-1} \\ 0 & b^{-1} & 0 \\ 0 & 0 & d^{-1} \end{bmatrix} \begin{bmatrix} \mathbf{i} \\ \mathbf{j} \\ \mathbf{k} \end{bmatrix} = \begin{bmatrix} a^{-1}\mathbf{i} - c(ad)^{-1}\mathbf{k} \\ b^{-1}\mathbf{j} \\ d^{-1}\mathbf{k} \end{bmatrix}. \tag{A8}$$

Obviously, the new coordinate system is oblique: the new $x^2$ and $x^3$ axes are along with the old $\bar{x}^2$ and $\bar{x}^3$ axes, while the new $x^1$ axis is inclined with the $\bar{x}^1$ at an angle $\varphi = \arctan(cd^{-1})$ (Fig. 1). The coincidence of the new $x^3$ and the old $z$ axis is important since in the vortex problem near $H_{c2}$, any $z$-independent quantity $\eta$ is $x^3$-independent:

$$\frac{\partial \eta}{\partial x^3} = \beta_3^{\bar{j}}\frac{\partial \eta}{\partial \bar{x}^{\bar{j}}} = \beta_3^{\bar{3}}\frac{\partial \eta}{\partial \bar{x}^3} = 0. \tag{A9}$$

The tensor components $\mu^{ij}$ in $(x^1, x^2, x^3)$ relates to $\mu^{\bar{i}\bar{j}}$ through the transformation relations:

$$\mu^{ij} = \beta_{\bar{m}}^i \beta_{\bar{n}}^j \mu^{\bar{m}\bar{n}}. \tag{A10}$$

The isotropic $\mu^{ij}$ requires that $\beta_{\bar{m}}^i \beta_{\bar{n}}^j \mu^{\bar{m}\bar{n}} = \delta^{ij}$, in matrix form it becomes

$$\begin{bmatrix} a & 0 & 0 \\ 0 & b & 0 \\ c & 0 & d \end{bmatrix} \begin{bmatrix} \mu_{xx} & 0 & \mu_{xz} \\ 0 & \mu_2 & 0 \\ \mu_{xz} & 0 & \mu_{zz} \end{bmatrix} \begin{bmatrix} a & 0 & 0 \\ 0 & b & 0 \\ c & 0 & d \end{bmatrix}^T = \begin{bmatrix} a^2\mu_{xx} & 0 & ac\mu_{xx} + ad\mu_{xz} \\ 0 & b^2\mu_2 & 0 \\ ac\mu_{xx} + ad\mu_{xz} & 0 & c^2\mu_{xx} + 2cd\mu_{xz} + d^2\mu_{zz} \end{bmatrix} = \begin{bmatrix} 1 & 0 & 0 \\ 0 & 1 & 0 \\ 0 & 0 & 1 \end{bmatrix}, \tag{A11}$$

and one obtains

$$a = \mu_{xx}^{-1/2}, \quad b = \mu_2^{-1/2}, \quad c = -\mu_2^{1/2}\mu_{xx}^{-1/2}\mu_{xz}, \quad d = (\mu_2\mu_{xx})^{1/2}, \tag{A12}$$

where the invariance relations Eq. (A4) and $\mu_1\mu_2\mu_3 = 1$ (obviously from the definition of the dimensionless inverse mass) have been used.

The metric tensor $g_{ij}$ of the new frame is obtained using Eq. (A8),

$$[g_{ij}] = [\mathbf{g}_i \cdot \mathbf{g}_j] = \begin{bmatrix} \mu_{xx} + \mu_{xz}^2 \mu_{xx}^{-1} & 0 & \mu_2^{-1/2} \mu_{xx}^{-1} \mu_{xz} \\ 0 & \mu_2 & 0 \\ \mu_2^{-1/2} \mu_{xx}^{-1} \mu_{xz} & 0 & (\mu_2 \mu_{xx})^{-1} \end{bmatrix}. \tag{A13}$$

The constant $g_{ij}$ make the covariant derivatives reduce to the partial ones, and since $\det[g_{ij}] = 1$, the Levi-Civita tensor $e^{ijk}$ preservers its $0$, $\pm 1$ form when transformed from $(x, y, z)$ to $(x^1, x^2, x^3)$. In the new frame $(x^1, x^2, x^3)$ with constant $g_{ij}$ and $\det[g_{ij}] = 1$, Eqs. (5a)-(5d) can be written as

$$\Pi_i \Pi_i \Psi = (1 - \alpha^{ij} u_{ij}) \Psi - |\Psi|^2 \Psi, \tag{A14a}$$

$$e^{ikl} \frac{\partial H_l}{\partial x^k} = \mathrm{Re}(\Psi^* \Pi_i \Psi), \tag{A14b}$$

$$\lambda^{ijkl} \langle u_{kl} \rangle + \alpha^{ij} \langle |\Psi|^2 \rangle = 0, \tag{A14c}$$

$$\frac{\partial}{\partial x^j} (\lambda^{ijkl} u_{kl} + \alpha^{ij} |\Psi|^2) = 0, \tag{A14d}$$

where $\Pi_i = (i\kappa)^{-1} \frac{\partial}{\partial x^i} - A_i$, and the covariant component of the magnetic field $H_l = g_{lm} H^m$ with the contravariant component

$$H^m = e^{mnr} \frac{\partial A_r}{\partial x^n}. \tag{A15}$$

$\Pi_i$, $\alpha^{ij}$, $u_{ij}$, $H_l$ and $\lambda^{ijkl}$ are obtained from their counterparts in the frame $(x, y, z)$ through the coordinate transformation relations: $\Pi_i = \beta_i^{\bar{j}} \Pi_{\bar{j}}$, $\alpha^{ij} = \beta_{\bar{k}}^i \beta_{\bar{l}}^j \alpha^{\bar{k}\bar{l}}$, $u_{ij} = \beta_i^{\bar{k}} \beta_j^{\bar{l}} u_{\bar{k}\bar{l}}$, $H_l = \beta_l^{\bar{m}} H_{\bar{m}}$ and $\lambda^{ijkl} = \beta_{\bar{q}}^i \beta_{\bar{r}}^j \beta_{\bar{s}}^k \beta_{\bar{t}}^l \lambda^{\bar{q}\bar{r}\bar{s}\bar{t}}$. As an example, we figure out the contravariant and covariant components of the magnetic field:

$$[H^i] = [\beta_{\bar{j}}^i H^{\bar{j}}] = \begin{bmatrix} a & 0 & 0 \\ 0 & b & 0 \\ c & 0 & d \end{bmatrix} \begin{bmatrix} H^{\bar{1}} \\ H^{\bar{2}} \\ H^{\bar{3}} \end{bmatrix} = \begin{bmatrix} \mu_{xx}^{-1/2} H_x \\ \mu_2^{-1/2} H_y \\ -\mu_2^{1/2} \mu_{xx}^{-1/2} \mu_{xz} H_x + \mu_2^{1/2} \mu_{xx}^{1/2} H_z \end{bmatrix},$$

$$[H_i] = [\beta_i^{\bar{j}} H_{\bar{j}}] = \begin{bmatrix} a^{-1} & 0 & -c(ad)^{-1} \\ 0 & b^{-1} & 0 \\ 0 & 0 & d^{-1} \end{bmatrix} \begin{bmatrix} H_{\bar{1}} \\ H_{\bar{2}} \\ H_{\bar{3}} \end{bmatrix} = \begin{bmatrix} \mu_{xx}^{1/2} H_x + \mu_{xx}^{-1/2} \mu_{xz} H_z \\ \mu_1^{1/2} H_y \\ \mu_1^{-1/2} \mu_{xx}^{-1/2} H_z \end{bmatrix}. \tag{A16}$$

**Appendix B: Derivations of Abrikosov identities in anisotropic deformable superconductors**

Introducing the operators $\Pi^+ = \Pi_1 + i\Pi_2$ and $\Pi^- = \Pi_1 - i\Pi_2$, we are allowed to carry out the following calculations:

$$\begin{aligned}\Pi^-\Pi^+ &= (\Pi_1 - i\Pi_2)(\Pi_1 + i\Pi_2)\\ &= \Pi_1^2 + \Pi_2^2 + i(\Pi_1\Pi_2 - \Pi_2\Pi_1)\\ &= \Pi_1^2 + \Pi_1^2 + \kappa^{-1}(\frac{\partial A_2}{\partial x^1} - \frac{\partial A_1}{\partial x^2})\\ &= \Pi_1^2 + \Pi_1^2 - \kappa^{-1}H^3 \quad,\end{aligned} \tag{A17}$$

where $H^3 = -(\frac{\partial A_2}{\partial x^1} - \frac{\partial A_1}{\partial x^2})$ [Eq. (A15)] has been used in the last equality. If we express the order parameter as $\Psi = \sqrt{\omega}\exp(i\chi)$ (as usual), then [bearing in mind that the module $\sqrt{\omega}$ of $\Psi$ is independent with $x^3$ according to Eq. (A9)]

$$\begin{aligned}\frac{\partial \Psi}{\partial x^3} &= \frac{\partial}{\partial x^3}[\sqrt{\omega}\exp(i\chi)]\\ &= \exp(i\chi)\frac{\partial \sqrt{\omega}}{\partial x^3} + i\sqrt{\omega}\exp(i\chi)\frac{\partial \chi}{\partial x^3}\\ &= i\Psi\frac{\partial \chi}{\partial x^3} \quad,\end{aligned} \tag{A18}$$

$$\begin{aligned}\Pi_3\Psi &= \left[(i\kappa)^{-1}\frac{\partial}{\partial x^3} - A_3\right]\Psi\\ &= \left[(i\kappa)^{-1}\frac{\partial \Psi}{\partial x^3} - A_3\Psi\right]\\ &= \left[\kappa^{-1}\frac{\partial \chi}{\partial x^3} - A_3\right]\Psi = Q_3\Psi \quad.\end{aligned} \tag{A19}$$

Here, we have introduced the gauge-invariant supermomentum $Q_i$,

$$Q_i = \kappa^{-1}\frac{\partial \chi}{\partial x^i} - A_i. \tag{A20}$$

Then, $\Pi_3^2\Psi$ reads

$$\begin{aligned}\Pi_3^2\Psi &= [(i\kappa)^{-1}\frac{\partial}{\partial x^3} - A_3]^2\Psi\\ &= [(i\kappa)^{-1}\frac{\partial}{\partial x^3} - A_3](Q_3\Psi)\\ &= (i\kappa)^{-1}\frac{\partial(Q_3\Psi)}{\partial x^3} - A_3(Q_3\Psi)\\ &= \kappa^{-1}Q_3\Psi\frac{\partial \chi}{\partial x^3} - A_3Q_3\Psi\\ &= Q_3^2\Psi \quad,\end{aligned} \tag{A21}$$

where $\partial Q_3/\partial x^3 = 0$ [Eq. (A9)] has been used. Thus, taking together Eqs. (A17)-(A21), $\Pi_i\Pi_i\Psi$ in Eq. (6a) yields

$$\Pi_i\Pi_i\Psi = \left(\Pi_1^2 + \Pi_2^2 + \Pi_3^2\right)\Psi = \left(\Pi^-\Pi^+ + \kappa^{-1}H^3 + Q_3^2\right)\Psi. \tag{A22}$$

Substituting Eq. (A22) into Eq. (6a), we obtain

$$\Pi^-\Pi^+\Psi = (1 - \alpha^{ij}u_{ij} - \kappa^{-1}H^3 - |\Psi|^2 - Q_3^2)\Psi. \tag{A23}$$